\documentclass[pra,twocolumn,superscriptaddress,floatfix,showpacs,10pt,letterpaper]{revtex4}
\usepackage{graphicx}
\usepackage{subfigure}
\usepackage{amssymb}
\usepackage{verbatim}
\usepackage{amsmath}
\usepackage{amscd}
\usepackage{amssymb}

\newcommand{\ket}[1]{|{#1}\rangle}

\newcommand{\ncd}{\newcommand}
\ncd{\QC}{$\mbox{QC}_{\cal{C}}\;$}
\ncd{\QCpr}{${\mbox{QC}_{\cal{C}}}^\prime\;$}
\ncd{\QCns}{$\mbox{QC}_{\cal{C}}$}
\ncd{\QCprns}{${\mbox{QC}_{\cal{C}}}^\prime$}
\ncd{\cskN}{{|\phi_{\{\kappa\} } \rangle}_{{\cal{C}}_N}}
\ncd{\cskNpr}{{|\phi_{\{\kappa^\prime\} } \rangle}_{{\cal{C}}_N}}
\ncd{\cskNtil}{{|\phi_{\{\tilde{\kappa} \} } \rangle}_{{\cal{C}}_N}}
\ncd{\csk}{{|\phi_{\{\kappa\} } \rangle}_{\cal{C}}}
\ncd{\csktil}{{|\phi_{\{\tilde{\kappa} \} } \rangle}_{\cal{C}}}
\ncd{\cskf}{|\phi_{\{\kappa\} } \rangle_{\cal{C}}}
\ncd{\csktilf}{|\phi_{\{\tilde{\kappa} \} } \rangle_{\cal{C}}}
\ncd{\bracsk}{\mbox{}_{\cal{C}}\langle\phi_{\{\kappa\} }|}
\ncd{\bracsktil}{\mbox{}_{\cal{C}}\langle\phi_{\{\tilde{\kappa} \}
}|} \ncd{\nbracsk}{\mbox{}_{\cal{C}}\langle\phi_{\{\kappa\} }}
\ncd{\nbracsktil}{\mbox{}_{\cal{C}}\langle\phi_{\{\tilde{\kappa} \}
}} \ncd{\cs}{|\phi \rangle_{\cal{C}}\;} \ncd{\csns}{|\phi
\rangle_{\cal{C}}} \ncd{\nbgh}{\text{nbgh}} \ncd{\Sab}{S^{ab}}
\ncd{\Sba}{S^{ba}} \ncd{\ds}{\displaystyle} \ncd{\ovl}{\overline}

\newtheorem{conjecture}{Conjecture}
\newtheorem{fact}{Fact}
\newtheorem{definition}{Definition}

\newtheorem{lemma}{Lemma}
\newtheorem{theorem}{Theorem}

\newtheorem{proposition}{Proposition}

\begin{document}

\title{Semi-Clifford operations, structure of $\mathcal{C}_k$ hierarchy, and \\ gate complexity for fault-tolerant quantum computation}

\author{Bei Zeng} \affiliation{Department of Physics, Massachusetts Institute of Technology, Cambridge, MA 02139, USA}

\author{Xie Chen} \affiliation{Department of Physics, Massachusetts Institute of Technology, Cambridge, MA 02139, USA}

\author{Isaac L. Chuang} \affiliation{Department of Physics, Massachusetts Institute of Technology, Cambridge, MA 02139, USA}

\date{\today}

\begin{abstract}

Teleportation is a crucial element in fault-tolerant quantum
computation and a complete understanding of its capacity is very
important for the practical implementation of optimal fault-tolerant
architectures. It is known that stabilizer codes support a natural
set of gates that can be more easily implemented by teleportation
than any other gates. These gates belong to the so called
$\mathcal{C}_k$ hierarchy introduced by Gottesman and Chuang (Nature
\textbf{402}, 390). Moreover, a subset of $\mathcal{C}_k$ gates,
called semi-Clifford operations, can be implemented by an even
simpler architecture than the traditional teleportation setup (Phys.
Rev. \textbf{A62}, 052316). However, the precise set of gates in
$\mathcal{C}_k$ remains unknown, even for a fixed number of qubits
$n$, which prevents us from knowing exactly what teleportation is
capable of. In this paper we study the structure of $\mathcal{C}_k$
in terms of semi-Clifford operations, which send by conjugation at least one maximal 
abelian subgroup of the $n$-qubit Pauli group into another one.
We show that for $n=1,2$, all
the $\mathcal{C}_k$ gates are semi-Clifford, which is also true for
$\{n=3,k=3\}$. However, this is no longer true for $\{n>2,k>3\}$. To
measure the capability of this teleportation primitive, we introduce
a quantity called `teleportation depth', which characterizes how
many teleportation steps are necessary, on average, to implement a
given gate. We calculate upper bounds for teleportation depth by
decomposing gates into both semi-Clifford $\mathcal{C}_k$ gates and
those $\mathcal{C}_k$ gates beyond semi-Clifford operations, and
compare their efficiency.

\end{abstract}

\pacs{03.67.Pp, 03.67.Lx} \maketitle 


\section{Introduction}

The discovery of quantum error-correcting codes and the theory of
fault-tolerant quantum computation have greatly improved the
long-term prospects for quantum computing technology
\cite{Nielsen,Preskill}. To implement fault-tolerant quantum
computation for a given quantum error-correcting code, protocols for
performing fault-tolerant operations are needed. The basic design
principle of a fault-tolerant operation protocol is that if only one
component in the procedure fails, then the failure causes at most
one error in each encoded block of qubits output from the procedure.

The most straightforward protocol is to use transversal gates
whenever possible. A transversal operation has the virtue that an
error occurring on the $k$th qubit in a block can only ever
propagate to the $k$th qubit of other blocks of the code, no matter
what other sequence of gates we perform before a complete
error-correction procedure \cite{Shor,Gottesman}. Unfortunately, it
is widely believed in the quantum information science community that
there does not exist a quantum error correcting code, upon which we
can perform universal quantum computations using just transversal
gates \cite{Gottesman}, and recently this belief is proved
\cite{ZCC}.

We therefore have to resort to other techniques, for instance
quantum teleportation \cite{Got} or state distillation
\cite{BravyiDistill}. The $\mathcal{C}_k$ hierarchy is introduced by
Gottesman and Chuang to implement fault-tolerant quantum computation
via teleportation \cite{Got}. The starting point is, if we can
perform the Pauli operations and measurements fault-tolerantly, we
can then perform all Clifford group operations fault-tolerantly by
teleportation. We can then use a similar technique to boot-strap the
way to universal fault-tolerant computation, using teleportation,
which gives a $\mathcal{C}_k$ hierarchy of quantum teleportation, as
defined below:

\begin{definition}
The sets $\mathcal{C}_k$ are defined in a recursive way as sets of
unitary operations $U$ that satisfy:
\begin{equation}
\mathcal{C}_{k+1}=\{U|U\mathcal{C}_1U^{\dagger}\subseteq\mathcal{C}_k\},
\end{equation}
where $\mathcal{C}_1$ is the Pauli group. We call a unitary
operation an $n$-qubit $\mathcal{C}_k$ gate if it belongs to the set
$\mathcal{C}_k$ and acts nontrivially on at most $n$ qubits.
\end{definition}

Note by definition $\mathcal{C}_2$ is the Clifford group, which
takes the Pauli group into itself. And
$\mathcal{C}_{k}\supset\mathcal{C}_{k-1}$, but $\mathcal{C}_k$ for
$k\geq 3$ is no longer a group.

\begin{figure}[htbp]
\includegraphics[width=3.00in]{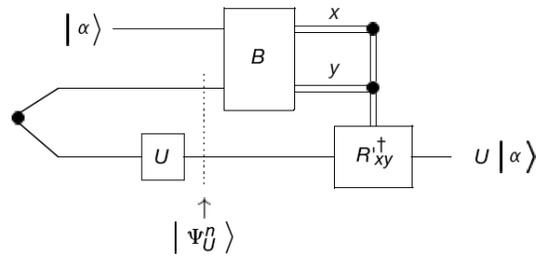}\caption{Two-bit teleportation scheme. $<$ denotes an \textbf{EPR}
pair, $B$ represents Bell-basis measurement, $R_{xy}' = U R_{xy}
U^{\dagger}$, where $R_{xy}$ is a Pauli operator. The double wires
carry classical bits and single wire carries qubits. Any gate in the
$\mathcal{C}_k$ hierarchy can be implemented fault-tolerantly using
this teleporation scheme.}\label{2bit}
\end{figure}

\begin{figure}[htbp]
\includegraphics[width=3.00in]{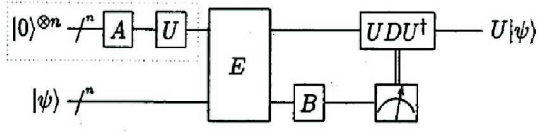}
\caption{One-bit teleportation scheme. For $Z$-teleportation, $A=I$,
$B=H$, $D=Z$, and $E$ is a CNOT gate with the first qubit as its
target. For $X$-teleportation, $A=H$, $B=I$, $D=X$, and $E$ is a
CNOT gate with the first qubit as its control. All semi-Clifford
$\mathcal{C}_k$ gates can be implemented fault-tolerantly using this
scheme.}\label{1bit}
\end{figure}

All the gates in $\mathcal{C}_k$ can be performed with the two-bit
teleportation scheme (FIG. \ref{2bit}) in a fault-tolerant manner.
Because, as proved in \cite{Gottesman}, it is possible to
fault-tolerantly prepare the ancilla state $|\Psi_U^n\rangle$, apply the
classically controlled correction operation $R_{xy}^{'\dagger}$, and
measure in Bell basis on a stabilizer code. However the precise set
of gates which form $\mathcal{C}_k$ is unknown, even for a fixed
number of qubits. It is demonstrated in \cite{Xinlan} that a subset
of $\mathcal{C}_k$ gates could be implemented by a different
architecture than the standard teleportation, called one-bit
teleportation, as shown in FIG. \ref{1bit}. Those gates adopt the
form $L_1VL_2$, where $V$ is a diagonal gate in $\mathcal{C}_k$ and
$L_1,L_2$ are two Clifford operations. Gates of this form are
recently studied in literature and are called the semi-Clifford
operations \cite{Gross}. In the following we will denote the
$n$-qubit Pauli group as $\mathcal{P}_n$ and a semi-Clifford
operation is defined to be a gate which sends at least one maximal
abelian subgroup of $\mathcal{P}_n$ to another maximal abelian one
under conjugation.

Due to the fact that one-bit teleportation needs only half the
number of ancilla qubits per teleportation than the standard two-bit
teleportation, it is important to understand the difference of
capabilities between one and two-bit teleportation for the practical
implementations of fault-tolerant architecture. It is conjectured in
\cite{Xinlan} that those two capabilities coincide for
$\{n=2,k=3\}$, which means that all the $\mathcal{C}_3$ gates for
two qubits are semi-Clifford operations.

\begin{figure}[htbp]
\includegraphics[width=3.00in]{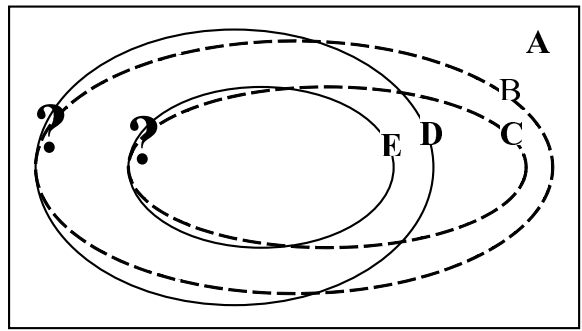}
\caption{Semi-Clifford operations versus $\mathcal{C}_k$ gates. A:
all gates; B: generalized semi-Clifford gates; C: semi-Clifford
gates; D: $\mathcal{C}_k$ gates; E: $\mathcal{C}_3$ gates. C is
strictly contained in B and E is strictly contained in D. The two
question marks indicate two open problems we have: whether $D$ is a
subset of $B$; and whether $E$ is a subset of $C$.} \label{Cksemi}
\end{figure}

In this paper, we prove this conjecture for a more general situation
where $\{n=1,2,\forall k\}$, and $\{n=3,k=3\}$. We then disprove it
for parameters $\{n>2,k>3\}$ by explicit construction of
counterexamples. We leave open the question for the parameters
$\{n>2,k=3\}$, and a more general problem of fully characterizing
the structure of $\mathcal{C}_k$: we conjecture that all gates in
$\mathcal{C}_k$ are something we refer to as generalized
semi-Clifford operations, i.e. a natural generalization of the
concept of semi-Clifford operation to the case including classical
permutations. Our results about this semi-Clifford operations versus
$\mathcal{C}_k$ gates relation can be visualized in FIG.
\ref{Cksemi}.

Just as in the usual circuit model, different gates are implemented
with different levels of complexity using this teleportation scheme.
It is then natural to ask the questions, how to characterize this
concept of gate complexity with concrete physical quantities, how
does this measure based on teleportation schemes compare with the
usual circuit depth, and what it implies for the practical
construction of quantum computation architecture. To answer these
questions, we introduce a quantity as a measure of gate complexity
for fault-tolerant quantum computation based on the $\mathcal{C}_k$
hierarchy, called the teleportation depth, which characterizes how
many teleportation steps are necessary, on average, to implement a
given gate. We demonstrate the effect of the existence of non
semi-Clifford operations in $\mathcal{C}_k$ on the estimation of the
upper bound for the teleportation depth, as well as some
quantitative difference between the capabilities of one and two-bit
teleportation.

The paper is organized as follows: Section II gives definition and
basic properties of semi-Clifford operations and generalized
semi-Clifford operations; in Section III we study the structure of
$\mathcal{C}_k$ hierarchy in terms of semi-Clifford and generalized
semi-Clifford operations; Section IV is devoted to the discussion of
teleportation depth and how it depends on the structure of
$\mathcal{C}_k$; and with Section V, we conclude our paper.

\section{Semi-Clifford operations and its generalization}

The concept of semi-Clifford operations was first introduced in
\cite{Gross}, to characterize the property of gates transforming
Pauli matrices acting on a single qubit. Here we generalize it to
the $n$-qubit case, through the following definition.

\begin{definition}
An $n$-qubit unitary operation is called semi-Clifford if it sends
by conjugation at least one maximal abelian subgroup of
$\mathcal{P}_n$ to another maximal abelian subgroup of
$\mathcal{P}_n$.

That is, if $U$ is an $n$-qubit semi-Clifford operation, then there
must exist at least one maximal abelian subgroup $G$ of
$\mathcal{P}_n$, such that $UGU^{\dagger}$ is another maximal
abelian subgroup of $\mathcal{P}_n$.
\end{definition}

The most basic property of a semi-Clifford operation is,
\begin{proposition}
If $R$ is a semi-Clifford operation, then there exist Clifford
operations $L_1,L_2$ such that $L_1RL_2$ is diagonal.
\end{proposition}
\textbf{Proof}: $Z_i$ represents the Pauli $Z$ operation on the
$i^{th}$ qubit. If  $R$ is an $n$-qubit semi-Clifford operation,
then there must exist $n$-qubit operations $L_1,L_2\in\mathcal{C}_2$
such that $RL_2Z_iL_2^{\dagger}R^{\dagger}=L_1^{\dagger}Z_iL_1$,
\cite{Gotthesis}, i.e.
$L_1RL_2Z_iL_2^{\dagger}R^{\dagger}L_1^{\dagger}=Z_i$ holds for any
$i=1...n$. Therefore, $(L_1RL_2)Z_i=Z_i(L_1RL_2)$, i.e. the
$n$-qubit gate $L_1RL_2$ is diagonal.$\square$

In other words, semi-Clifford operations are those gates
diagonalizable `up to Clifford multiplications'. Thus the structure
problem of the whole set of semi-Clifford operations is reduced to
that of the diagonal subset within it.

As we shall see later, the notion of semi-Clifford operations is
useful in characterizing some but not all gates in the
$\mathcal{C}_k$ hierarchy. More generally, we might also consider
those gates with properties of transforming the span, or in other
words the group algebra over the complex field, of a maximal abelian
subgroup of $\mathcal{P}_n$.

\begin{definition}
A generalized semi-Clifford operation on $n$ qubits is defined to
send by conjugation the span of at least one maximal abelian
subgroup of $\mathcal{P}_n$ to the span of another maximal abelian
subgroup of $\mathcal{P}_n$.
\end{definition}

Denote $\langle S_i\rangle$ the group generated by a set of operators $\{S_i\}$, and denote 
the span of the group $\langle S_i\rangle$ as
$\mathfrak{C}(\langle S_i\rangle)$. Then in a more mathmatical form we can write
the above definition as:

If $U$ is a generalized semi-Clifford operation on $n$ qubits, then
there must exist at least one maximal abelian subgroup $G=\langle g_i\rangle$ of
$\mathcal{P}_n$, such that for all $s\in \mathfrak{C}(\langle g_i\rangle)$, 
$UsU^{\dagger}\in\mathfrak{C}(U\langle g_i\rangle U^{\dagger})$, where $UGU^{\dagger}$ is another maximal abelian subgroup of $\mathcal{P}_n$.

Then the basic property of a generalized semi-Clifford operation is,
\begin{proposition}
If $R$ is a generalized semi-Clifford operation, then there exist
Clifford operations $L_1,L_2$, and a classical permutation operator
$P$ such that $PL_1RL_2$ is diagonal.
\end{proposition}
\textbf{Proof}: If  $R$ is a generalized semi-Clifford operation,
then there must exist $L_1,L_2\in\mathcal{C}_2$ such that
$RL_2\mathfrak{C}(\langle
Z_i\rangle_{i=1}^n)L_2^{\dagger}R^{\dagger}=L_1^{\dagger}\mathfrak{C}(\langle
Z_i\rangle_{i=1}^n)L_1$, i.e. $L_1RL_2\mathfrak{C}(\langle
Z_i\rangle_{i=1}^n)L_2^{\dagger}R^{\dagger}L_1^{\dagger}=\mathfrak{C}(\langle
Z_i\rangle_{i=1}^n)$. That is, $L_1RL_2$ maps all the diagonal
matrices to diagonal matrices, therefore $L_1RL_2$ must be a
monomial matrices, i.e. there exist a permutation matrix $P$ and a
diagonal matrix $V$, such that $L_1RL_2=P^{\dagger}V\Rightarrow
PL_1RL_2$ is diagonal.$\square$

Note for the single qubit case, i.e. $n=1$, the concepts of
semi-Clifford operation and generalized semi-Clifford operation
coincide.

\section{The structure of $\mathcal{C}_k$}

In this section we study the structure of gates in $\mathcal{C}_k$.
To begin with, we study some basic properties of $\mathcal{C}_k$
gates. Then we give our main results as structure theorems, which
state that all the $\mathcal{C}_k$ gates are semi-Clifford when
$\{n=1,2,\forall k\}$ and $\{n=3,k=3\}$, but for $\{n>2,k>3\}$ there
are examples of $\mathcal{C}_k$ gates which are non-semi-Clifford.
We then discuss the open question for the parameters $\{n>2,k=3\}$,
and based on the constructed counterexamples we conjecture that all
$\mathcal{C}_k$ gates are generalized semi-Clifford operations.

It should be noted that the set of $n$-qubit $\mathcal{C}_k$ gates
is always strictly contained in the set of $n$-qubit
$\mathcal{C}_{k+1}$ gates. In \cite{Got}, explicit examples are
given to support this statement. If we denote as $\Lambda_{n-1}(U)$
the $n$-qubit gate which applies $U$ to the $n$th qubit only if the
first $n-1$ qubits are all in the state $\ket{1}$, then
$\Lambda_{n-1}(\text{diag}(1,e^{2\pi/2^m}))$ is in
$\mathcal{C}_{m+n-1}\setminus\mathcal{C}_{m+n-2}$.

\subsection{Basic properties}

We first state an important property of gates in $\mathcal{C}_k$,
which reduce the problem of characterizing the structure of
$\mathcal{C}_k$ into a problem of characterizing a certain subset of
gates in $\mathcal{C}_k$.

\begin{proposition}
If $R\in \mathcal{C}_k$, then $L_1RL_2\in \mathcal{C}_k$, where
$L_1,L_2\in \mathcal{C}_2$, $k\geq 2$.
\end{proposition}
\textbf{Proof}: We prove this proposition by induction.

i) It is obviously true for $k=2$;

ii) Assume it is true for $k$;

iii) For $k+1$, $R\in \mathcal{C}_{k+1}$ implies $RAR^{\dagger}\in
\mathcal{C}_{k}$, where $A\in \mathcal{C}_{1}$. If we conjugate $A$
by $L_1RL_2$, we get
\begin{equation}
L_1RL_2A(L_1RL_2)^{\dagger}=L_1R(L_2AL_2^{\dagger})R^{\dagger}L_1^{\dagger}.
\end{equation}

Since $L_1,L_2\in \mathcal{C}_2$, $L_1^{\dagger},L_2^{\dagger}$ are
in $\mathcal{C}_2$ also. And because $L_2AL_2^{\dagger} \in
\mathcal{C}_1$, $R(L_2AL_2^{\dagger})R^{\dagger} \in \mathcal{C}_k$.
According to assumption ii),
$L_1R(L_2AL_2^{\dagger})R^{\dagger}L_1^{\dagger} \in \mathcal{C}_k$.
Finally as we can see from Eqn (2), $L_1RL_2 \in
\mathcal{C}_{k+1}$.$\square$

According to Proposition 3, in order to characterize the full
structure of $\mathcal{C}_k$, we only need to characterize the
structure of a subset of it which generates the whole set with
Clifford multiplications.

It is known that $\mathcal{C}_k$ is not a group for $k>2$ and its
structure is in general hard to characterize. However, if we denote
all the diagonal gates in $\mathcal{C}_k$ as $\mathcal{F}_k$, then
we have the following:
\begin{proposition}
$\mathcal{F}_{k}$ is a group.
\end{proposition}

If we can characterize the group structure of $\mathcal{F}_k$, then
the structure of the $\mathcal{C}_k$ subset $\{L_1F_kL_2\}$ is known
to us ($L_1, L_2 \in \mathcal{C}_2$, $F_k \in \mathcal{F}_k$).
According to Proposition 1, this is just the set of all
semi-Clifford operations in $\mathcal{C}_k$. In the next section, we
will repeatedly use this fact to gain knowledge about semi-Clifford
$\mathcal{C}_k$ gates from the group structure of $\mathcal{F}_k$
and for now we will give a brief proof of the above proposition.

\textbf{Proof}: We prove by induction.

i) It is of course true for $k=2$;

ii) Assume it is true for $k$, i.e. $\mathcal{F}_{k}$ is group;

iii) Then for $k+1$, note for any $F_{k+1} \in\mathcal{F}_{k+1}$,
$F_{k+1}MF_{k+1}^{\dagger}=F_{k}M=MF'_{k}$, for non-diagonal
$M\in\mathcal{C}_1$, where $F_{k}, F'_{k}\in \mathcal{F}_{k}$.

a) If $F_{k+1} \in\mathcal{F}_{k+1}$, then $F_{k+1}^{\dagger}
\in\mathcal{F}_{k+1}$, since
$F_{k+1}^{\dagger}MF_{k+1}=F_{k}^{\dagger}M=MF^{'\dagger}_{k}$,
which is in $\mathcal{F}_{k}$ by assumption ii).

b) If $F_{1k},F_{2k}\in\mathcal{F}_k$, then
$F_{1k}F_{2k}\in\mathcal{F}_k$, since
$F_{1k-1}F_{2k-1}\in\mathcal{F}_{k-1}$.$\square$

According to this proposition, all semi-Clifford $\mathcal{C}_k$
gates can be characterized by the group structure of diagonal
$\mathcal{C}_k$ gates.

\subsection{Structure theorems}

Our main results about the structure of $\mathcal{C}_k$ are the
following three theorems, which state that all the $\mathcal{C}_k$
gates are semi-Clifford when $\{n=1,2,\forall k\}$ and
$\{n=3,k=3\}$, but it is no longer true for $\{n>2,k>3\}$.

\begin{theorem}
All gates in $\mathcal{C}_k$ are semi-Clifford operations for
$(n=1,2,\forall k)$.
\end{theorem}
\textbf{Proof}: Here we prove the case of $n=2$. The proof of the
$n=1$ case is similar but can also be checked by direct calculation
and lead to a complete classification of all 1-qubit $\mathcal{C}_k$
gates according to the group structure of diagonal 1-qubit
$\mathcal{C}_k$ gates. We give details for the $n=1$ case in
appendix.

For $n=2$, we prove this theorem by induction:

i) It is obviously true for $k=1,2$;

ii) Assume it is true for $k$;

iii) For $k+1$:

a) We calculate the set $S_1=\{L_1V\}$ for all
$L_1\in\mathcal{C}_2$, where $V\in\mathcal{F}_k$. Note by assumption
ii), $S_1$ gives us all the elements in $\mathcal{C}_k$ up to
Clifford conjugation.

b) Note in general
$V=\text{diag}\{e^{i\alpha},e^{i\beta},e^{i\gamma},e^{i\delta}\}$
for some angles $\alpha,\beta,\gamma$ and $\delta$. By exhaustive
calculation with all $L_1\in\mathcal{C}_2$ we show that if there
exists an element $V_s\in S_1$ such that $V_s$ is trace zero and
Hermitian, then
$V=\text{diag}\{e^{-i\theta_1},e^{-i\theta_2},e^{i\theta_2},e^{i\theta_1}\}$
for some $\theta_1$ and $\theta_2$. Furthermore, we can again show
by  exhaustive calculation with all $L_1\in\mathcal{C}_2$ that the
only trace zero and Hermitian $V_s\in S_1$ is of the following form
up to Clifford conjugation:
\begin{equation}
V_s=\left(\begin{matrix}
0 & 0 & 0 &e^{-i\theta_1} \\
0 & 0 &e^{-i\theta_2} & 0\\
0 & e^{i\theta_2} & 0 & 0\\
e^{i\theta_1} & 0 & 0 & 0
\end{matrix}\right).
\end{equation}

c) We calculate the set $S_2=\{L_1V_sL_1^{\dagger}\}$ for all
$L_1\in\mathcal{C}_2$, which by assumption ii) and fact b) gives all
the elements in $\mathcal{C}_k$ which are trace zero and Hermitian.

d) We show that for any two-qubit gate $U$ such that
$UV_sU^{\dagger}=Z_1$ and $\{U\mathcal{P}_2U^{\dagger}\}\subseteq
S_2$, there exist $L_1,L_2\in\mathcal{C}_2$ such that $L_1UL_2$ is
diagonal. This can be started from studying the eigenvectors of
$V_s$, which can be chosen of the form

\begin{equation}
U=\frac{1}{\sqrt{2}}\left(\begin{matrix}
1 & 0 & 0 &1 \\
0 & 1 &1 & 0\\
0 & e^{i\theta_2} & -e^{i\theta_2} & 0\\
e^{i\theta_1} & 0 & 0 & -e^{i\theta_1}
\end{matrix}\right),
\end{equation}
and carefully considering the possible phase of each eigenvector and
the possible superposition of the eigenvectors due to the degeneracy
of the eigenvalues, similar to the process shown in Appendix
A.$\square$

\begin{theorem}
All gates in $\mathcal{C}_k$ are semi-Clifford operations for
$\{n=3,k=3\}$.
\end{theorem}
\textbf{Proof}: We prove this theorem exhaustively using the
following proposition:
\begin{proposition}
An $n$-qubit $\mathcal{C}_k$ gate $U$ is semi-Clifford if and only
if the group $\{U\mathcal{P}_nU^{\dagger}\}\cap\mathcal{P}_n$
contains a maximally abelian subgroup of $\mathcal{P}_n$.
\end{proposition}
\textbf{Proof}: Suppose $U=L_1VL_2$, then
$U\mathcal{P}_nU^{\dagger}=L_1VL_2\mathcal{P}_nL_1^{\dagger}V^{\dagger}L_1^{\dagger}=L_1V\mathcal{P}_nV^{\dagger}L_1^{\dagger}\supset\{L_1Z_iL_1^{\dagger}\}_{i=1}^n$.

On the contrary, if $\{U\mathcal{P}_nU^{\dagger}\}\cap\mathcal{P}_n$
contains a maximal abelian subgroup of $\mathcal{P}_n$, then there
must exist $L_1,L_2 \in \mathcal{C}_2$ such that
$UL_1^{\dagger}Z_iL_1U^{\dagger}=L_2Z_iL_2^{\dagger}$, i.e.
$L_2^{\dagger}UL_1^{\dagger}Z_iL_1U^{\dagger}L_2=Z_i$ holds for any
$i=1...n$. Therefore,
$(L_2^{\dagger}UL_1^{\dagger})Z_i=Z_i(L_2^{\dagger}UL_1^{\dagger})$,
$\Rightarrow L_2^{\dagger}UL_1^{\dagger}$ is diagonal. If we denote
this diagonal gate as $V$, $L_2^{\dagger}UL_1^{\dagger}=V\Rightarrow
U=L_1VL_2$.

Therefore, by exhaustive study with the subgroups of the three-qubit
Clifford group which are isomorphic to $\mathcal{P}_3$, we complete
the proof of this theorem. More detailed analysis about this is
given in Appendix B. The calculation is done using GAP
\cite{GAP}.$\square$

\begin{figure}[htbp]
\includegraphics[width=2.00in]{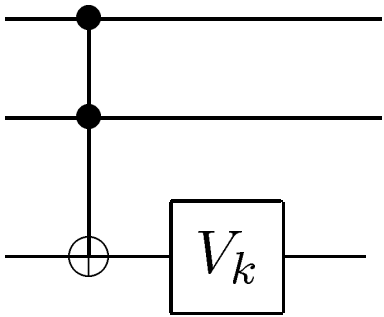}
\caption{A non-Clifford-diagonalizable $\mathcal{C}_k$ gate $W_k$.
$V_{k}=diag(1,e^{i\pi/2^{k-1}})$.}\label{fig4}
\end{figure}

\begin{theorem}
Not all gates in $\mathcal{C}_k$ are semi-Clifford operations for
$(n>2, k>3)$.
\end{theorem}
\textbf{Proof}: Actually we only need to prove this theorem for
$n=3,k=4$ then it naturally holds for all the other parameters of
$\{n>2,k>3\}$. However we would like to explicitly construct
examples for all $\{n=3,k>4\}$. Define $W_{k}$ as in FIG. \ref{fig4}.

\begin{proposition}
The gate
\begin{equation}
W_k=T(c_1, c_2, t_3) \otimes V_{3,k}
\end{equation}
is a $\mathcal{C}_{k+1}$ operation but not a semi-Clifford
operation, where $T(c_1, c_2, t_3)$ is Toffoli gate with the 1st and
2nd qubit as its control and the 3rd qubit as its target, $V_{3,k}$
is single qubit operator $diag(1,exp(i\pi/2^{k-1}))$ on the 3rd
qubit.
\end{proposition}
\textbf{Proof}:

To prove that $W_k$ is in $\mathcal{C}_{k+1}$,

i) When $k=2$, $V_{k} = diag\{1, i\} \in \mathcal{C}_2$. $W_2$ is of
the form $LR$, where $L$ is a Clifford operation and $R$ is the
Toffoli gate. According to Proposition 3, $W_2$ and the Toffoli gate
are both in $\mathcal{C}_3$.

ii) For $k > 2$, direct calculation shows that
$\{W_kZ_iW_k^{\dagger}\} \subset \mathcal{C}_2$, $i = 1, 2, 3$.
$W_kX_1W_k^{\dagger} \in \mathcal{C}_k$, $W_kX_2W_k^{\dagger} \in
\mathcal{C}_k$, $W_kX_3W_k^{\dagger} \in \mathcal{C}_{k-1}$. The
images of $X_i$'s under the conjugation of $W_k$ can all be written
in the form $W_kX_iW_k^{\dagger} = X_i F_{ki} = F'_{ki} X_i$, where
$F_{k1}$, $F'_{k1}$, $F_{k2}$, $F'_{k2}$ are diagonal gates in
$\mathcal{C}_k$ and $F_{k3}$, $F'_{k3}$ are diagonal single qubit
gates in $\mathcal{C}_{k-1}$ acting on the third qubit.

The image of the whole 3-qubit Pauli group $\{W_k\mathcal{P}_3
W_k^{\dagger}\}$ is generated by the six elements shown above. As
multiplication by Clifford gates preserves the $\mathcal{C}_k$
hierarchy, we only need to check the images of Pauli operations
which are composed of two or more $X_i$'s and see if their images
are still in $\mathcal{C}_k$.

This is obviously true considering the special form of
$\{W_kX_iW_k^{\dagger}\}$. Multiplication of any two of them is of
the form $W_kX_iX_jW_k^{\dagger} = X_i F_{ki} F'_{kj} X_j$. This is
in $\mathcal{C}_k$ as the diagonal $\mathcal{C}_k$ gates form a
group. Further more, multiplication of all of them takes the form
$W_kX_1X_2X_3W_k^{\dagger} = X_1 F_{ki} F'_{kj} X_2 F'_{k3} X_3$. As
$F'_{k3}$ is a single qubit operation on the third qubit,
$W_kX_1X_2X_3W_k^{\dagger} =X_1 F_{ki} F'_{kj} F'_{k3} X_2 X_3$.
This is again a $\mathcal{C}_k$ gate because of the group structure
of diagonal $\mathcal{C}_k$ gates.

Therefore, we have checked explicitly that $W_k \in
\mathcal{C}_{k+1}$.

To prove that $W_k$ is not semi-Clifford, we can exhaustively
calcultate $\{W_k\mathcal{P}_3W_k^{\dagger}\}$ and find its
intersection with $\mathcal{P}_3$. The fact that
$\{W_k\mathcal{P}_3W_k^{\dagger}\} \cap \mathcal{P}_3$ does not
contain a maximally abelian subgroup of $\mathcal{P}_3$ implies that
$W_k$ is not semi-Clifford, due to Proposition 5.

With this example we have directly proved Theorem 3. $\square$

\subsection{Open problems}

Let us try to understand more about the structure theorems we have
in the previous section.

First recall from \cite{Xinlan} that the controlled-Hadamard gate
$\Lambda_1(H)$, which is a $\mathcal{C}_3$ gate, is explicitly shown
to be semi-Clifford. We can also view this from the perspective of
Proposition 5, by noting that
$\Lambda_1(H)Z_1\Lambda_1(H)^{\dagger}=Z_1$,
$\Lambda_1(H)Y_2\Lambda_1(H)^{\dagger}=Z_1\otimes -Y_2$, which means
that the maximal abelian subgroup of the Pauli group generated by
$\langle Z_1,Y_2\rangle\times\langle\pm 1,\pm i\rangle$ is in the
image of $\Lambda_1(H)$. However, if we consider $W_3$ from the
perspective of Proposition 5, we get $W_3Z_1W_3^{\dagger}=Z_1$,
$W_3Z_2W_3^{\dagger}=Z_2$,
$W_3Z_3W_3^{\dagger}=\Lambda_1(Z_2)\otimes Z_3$. Note this do not
give us a maximal abelian subgroup of the Pauli group
$\langle Z_1,Z_2,Z_3\rangle\times\langle\pm 1,\pm i\rangle$, 
due to the effect of $\Lambda_1(Z_2)$ caused by
conjugating through the Toffoli gate. This intuitively explains why
Theorem 3 could be true, but no counterexample to Theorem 2 exists.

Note that $W_k$ is actually a generalized semi-Clifford operation,
which is apparent from its form. Also, the construction of the
series of gates $W_k$, as well as their extensions to $n>3$ qubits,
cannot give any non-semi-Clifford $\mathcal{C}_3$ gate. We then have
the following conjectures on the open problem of the structure of
$\mathcal{C}_k$ hierarchy in general.

\begin{conjecture}
All gates in $\mathcal{C}_3$ are semi-Clifford operations.
\end{conjecture}

\begin{conjecture}
All gates in $\mathcal{C}_k$ are generalized semi-Clifford operations.
\end{conjecture}

\section{The teleportation depth}

Teleportation, as a computational primitive, is a crucial element
providing universal quantum computation to fault-tolerant schemes
based on stabilizer codes. However, not all gates are of equal
complexity in this scheme. To actually incorporate this technique in
the construction of practical computational architecture, it is
useful to know which gates are easier to implement and which are
harder, so that we could achieve optimal efficiency in performing a
computational task. In the circuit model of quantum computation, we
face the same problem and in that case `circuit depth'
was introduced \cite{Yao} to characterize the number of simple one
and two-qubit gates needed to implement an operation. While this
provides a good measure of gate complexity, it does not take into
consideration of fault-tolerance. It is interesting to have measures
quantifying fault-tolerant gate complexity to be compared with
`circuit depth' to give us a better understanding of the
computational tasks at hand.

Based on the $\mathcal{C}_k$ hierarchy introduced in \cite{Got} and
the knowledge of its structure gained in previous section, we define
a measure of gate complexity for the teleportation protocol, called
the teleportation depth, which characterizes how many teleportation
steps are necessary, on average, to implement a given gate. Since
any teleportation unavoidably causes randomness, we need to figure
out a certain point to start with, i.e. we should assume in advance
that some kind of gates can be performed fault-tolerantly. We know
that a fault-tolerant protocol is usually associated with some
quantum error-correcting codes. Self-dual CSS codes, such as the
$7$-qubit Steane code, admit all gates in the Clifford group to be
transversal \cite{Gotthesis}. In such a situation, we only need to
teleport the gate outside the Clifford group, and in the following,
we will assume this as a starting point. The advantage of doing
this, in practice, is that due to Proposition 1, we have the freedom
of preparing the ancilla states up to some Clifford multiplications.

\subsection{Definition of the teleportation depth}

With the standard two-bit teleportation scheme (FIG. \ref{2bit}) in
mind, it is easy to see that all gates in the $\mathcal{C}_k$
hierarchy can be teleported fault-tolerantly as a whole in a
recursive manner. Suppose $U$ is an $n$-qubit $\mathcal{C}_k$ gate.
The ancilla state can be fault-tolerantly prepared and all the
elements in the teleportation circuit of $U$ are in $\mathcal{C}_2$
and can be performed fault-tolerantly, except the classically
controlled operation $U_1 = R_{xy}' = U R_{xy} U^{\dagger}$, where
$R_{xy}$ is an operator in $\mathcal{C}_1$ which depends on the
(random) Bell-basis measurement outcomes $xy$. However, as $U$ is in
$\mathcal{C}_k$, $U_1$ is in general a $\mathcal{C}_{k-1}$ operation
and can be implemented again by teleportation. In this way, after
each teleportation step, a $\mathcal{C}_k$ gate is mapped to another
gate one level lower. This recursive procedure terminates when $U_i$
is in $\mathcal{C}_2$.

Based on the above picture we give a more formal definition of
teleportation, which characterizes its randomness nature.

\begin{definition}
The teleportation map $f$ takes an $n$-qubit operator $A$ to a set
of operators via the following manner:
\begin{equation}
f: A\rightarrow \{AP_{j_1}A^{\dagger}\}_{j_1=1}^{4^{n+1}},
\end{equation}
where $P_i$ are elements of the $n$-qubit Pauli group
$\mathcal{P}_n$.
\end{definition}

Note
\begin{equation}
f\circ f: A\rightarrow
\{(AP_{j_1}A^{\dagger})P_{j_2}(AP_{j_1}A^{\dagger})^{\dagger}\}_{j_1,j_2=1}^{4^{n+1}},
\end{equation}
and
\begin{eqnarray}
f\circ f\circ f: A\rightarrow \nonumber\\
\{((AP_{j_1}A^{\dagger})P_{j_2}(AP_{j_1}A^{\dagger})^{\dagger})P_{j_3}((AP_{j_1}A^{\dagger})P_{j_2} \nonumber\\
(AP_{j_1}A^{\dagger})^{\dagger})^{\dagger}\}_{j_1,j_2,j_3=1}^{4^{n+1}},
\cdots
\end{eqnarray}

Each element of image of the map $f^{\circ m}$ on $A$ is associated
with a set
\begin{equation}
S=\{j_1,j_2\ldots,j_m\}.
\end{equation}

Denote $f^{\circ m}_S(A)$ as the element in image of the map
$f^{\circ m}$ on $A$ associated with the set $S$. Each element in
the image occurs with equal probability.

\begin{definition}
$f^{\circ m}_S(A)$ terminates if $f^{\circ m}_S(A)\in
\mathcal{C}_2$.
\end{definition}

If $f^{\circ m_1}_S(A)$ terminates, then $f^{\circ m_2}_{S'}(A)$
terminates for any $m_2\geq m_1$, and
$S'=\{j_1,j_2\ldots,j_{m_1},\ldots,j_{m_2}\}$. Therefore, for each
$f^{\circ m}_S(A)$ that terminates, there must exist a set $S_{min}$
with the minimal size such that $f^{\circ |S_{min}|}_{S_{min}}(A)$
terminates, where $S_{min}=\{j_1,j_2\ldots,j_{m'}\}$
($m'=|S_{min}|$). In our following discussions, we will only
consider sets $S$ which are minimal in this sense.

This mapping procedure works directly on $\mathcal{C}_k$ gates. If
$W$ is an $n$-qubit $\mathcal{C}_k$ gate, then there is no need to
decompose it into consecutive application of several other gates and
we say we can `direct teleport' $W$. $W$ is in $\mathcal{C}_k$ iff
$\forall S$, $f^{\circ (k-2)}_S(A) \in \mathcal{C}_2$, and $\exists
S'$, s.t. $f^{\circ (k-3)}_{S'}(A) \notin \mathcal{C}_2$.

Among all $\mathcal{C}_k$ gates, the set of semi-Clifford operations
have the special property that they can be teleported with only half
the ancilla resources as in a standard teleportation scheme. This
`one-bit teleportation scheme' is illustrated in FIG. \ref{1bit}.
This scheme also complies with the mapping description given above.
Instead of Bell basis measurement, randomness in one-bit
teleportation scheme comes from single qubit measurement and $P_j$
belongs to a maximal abelian subgroup of the whole $n$-qubit Pauli
group in general.

To teleport an arbitrary $n$-qubit gate $A$, we can first decompose
$A$ into the $\mathcal{C}_k$ hierarchy, $A = A_1 A_2 \dots A_r$,
where $A_i \in \mathcal{C}_{k_i}$, because we only know how to
teleport $\mathcal{C}_k$ gates fault-tolerantly. We call this
procedure `decomposition of $A$  into $\mathcal{C}_{\infty}$'.
Suppose that to teleport each gate $A_i$, $m_i$ maps are needed on
average, with average taken over all possible set $S = \{j_1,
j_2\dots, j_m\}$. Then the teleportation depth of $A$ is defined as
follows.

\begin{definition}
The teleportation depth of a gate $A$, denoted as $T$, is the
minimal sum of all $m_i$--the average number of teleportation steps
needed to implement each component gate of $A$--where the minimum is
taken over all possible decompositions of $A$ into
$\mathcal{C}_{\infty}$.\label{depth}
\end{definition}

Due to Definition \ref{depth}, in order to calculate the teleportation
depth of a given gate $A$, one needs to find all possible
decompositions of $A$ into $\mathcal{C}_{\infty}$ gates and
calculate the corresponding depth, then minimize over all of them.
This is generally intractable, but one may expect to upper bound the
depth with some particular decomposition of $A$ into
$\mathcal{C}_{\infty}$ gates.

Let us first consider the case of an $n$-qubit $\mathcal{C}_k$ gate.

\begin{definition}
$T(n,k)$ is the teleportation depth of an $n$-qubit $\mathcal{C}_k$
gate.
\end{definition}

As such a gate can be teleported directly, $T(n,k)$ is upper bounded
by the average number of steps needed in this direct teleportation
scheme to terminate the teleportation procedure.

\begin{equation}
T(n,k) \leq \frac{1}{N} \sum_S |S|
\end{equation}where the summation is over all possible (minimal) sets $S$ and $N$
is the number of such sets.

However, when $k\rightarrow\infty$, it is not obvious that the above
summation will converge. We will show that this is true. Then for an
arbitratry gate $A$, by decomposing $A$ into a finite series of
$\mathcal{C}_k$ gates, we can see that the teleportation depth of
$A$ turns out to be finite. Then we do not actually require the
procedure to terminate within a finite number of steps.

Different teleportation schemes, for example one-bit and two-bit
teleportation, give different upper bounds on teleportation depth
for a certain circuit. While for some circuits one scheme is
obviously more efficient than others, the comparison among different
schemes in other case may not be so straightforward and may depend
sensitively on various parameters in the circuit. In the following
sections, we study such dependence and present surprising results
beyond our usual expectation with examples from important quantum
circuits.

\subsection{Teleportation depth of semi-Clifford $\mathcal{C}_k$ gates}

We first calculate explicitly an upper bound for the teleportation
depth of semi-Clifford $n$-qubit $\mathcal{C}_k$ gates. We know from
\cite{Xinlan} that this kind of gate can be teleported directly with
the architecture of one-bit teleporation and we denote the upper
bound calculated with this `\textbf{one-bit}' `\textbf{direct}'
teleportation procedure as $T_1(n,k)$. For a general $n$-qubit gate,
if it is possible to decompose it into a series of semi-Clifford
$\mathcal{C}_k$ operations, the upper bound of teleportation depth
obtained by teleporting each part separately using one-bit
teleportation scheme is in general denoted as $T_1$.

\begin{definition}
$T_1$ is the average total number of teleportation steps needed to
teleport separately each semi-Clifford $\mathcal{C}_k$ component of
a quantum circuit using the one-bit teleportation scheme, if such a
decomposition is possible.

More specifically, $T_1(n,k)$ is the average number of teleportation
steps needed to teleport an $n$-qubit semi-Clifford $\mathcal{C}_k$
gate directly (i.e. without decomposition) using the one-bit
teleportation scheme.
\end{definition}

Apparently we have $T(n,k) \leq T_1(n,k)$ in general.

The probability that the teleportation process terminates
immediately after one teleportation step equals the percentage
weight of a maximal abelian subgroup in the whole Pauli group, which
is $\frac{1}{2^n}$ for an $n$-qubit Pauli group. Now each
teleportation step may have two possible endings: i) with
probability $p=\frac{1}{2^n}$, $\{U\mathcal{P}_nU^{\dagger}\} \in
\mathcal{P}_n$ and the process terminates; ii) with probability
$1-p$, $\{U\mathcal{P}_nU^{\dagger}\}$ is a general $n$-qubit
$\mathcal{C}_{k-1}$ gate and the process goes on. The upper bound of
teleportation depth calculated with this process is then

\begin{eqnarray}
T_1(n,k) &=& p\sum\limits_{s=1}^{k-3} s(1-p)^{s-1}+(k-2)(1-p)^{k-3}\nonumber\\
&=&2^n\left(1-(1-\frac{1}{2^n})^{k-2}\right).\label{depthck}
\end{eqnarray}

\begin{figure}[htbp]
\includegraphics[width=3.00in]{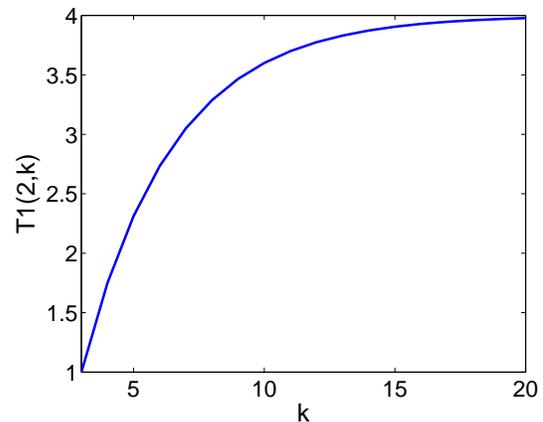}
\caption{The behavior of
$T_1(2,k)=4(1-(3/4)^{k-2})$}\label{depthtwoqubits}
\end{figure}

It is clearly seen from Eq. (\ref{depthck}) that $T_1(n,k)$
converges to $2^n$ when $k\rightarrow\infty$, which means that
$T(n,k)$ is in general bounded. For instance, when $n=2$, Eq.
(\ref{depthck}) tells us $T(2,k) \leq T_1(2,k) = 4(1-(3/4)^{k-2})$.
The behavior of $T_1(2,k)$ is shown in FIG. \ref{depthtwoqubits}.
However, since $T_1(2,k) = 4(1-(3/4)^{k-2}) \leq 4(1-(1/2)^{k-2}) =
2T_1(1,k)$, we find that teleporting two single-qubit semi-Clifford
$\mathcal{C}_k$ gates together using the one-bit teleportation
scheme needs fewer teleportation steps than to teleport each of them
separately.

Since $1-\frac{1}{2^n}<1$, $T_1(n,k)$ quickly reaches $2^n$ as $k$
grows. Therefore, generally, the upper bound of the teleportation
depth of a $\mathcal{C}_k$ gate given by `direct teleportation' is
not determined by $k$, but by the number of qubits $n$ it actually
acts on. Moreover, since $T_1(n,\infty)=2^n$, i.e. the upper bound
of teleportation depth increases exponentially with $n$, in
generally, when $n,k$ are large, it is better to decompose an
$n$-qubit $\mathcal{C}_k$ gate into some one and two-qubits gates to
get a lower upper bound.

However, if $k\sim P(n)$, where $P(n)$ is a polynomial in $n$, then
$T_1(n,k)$ scales as $P(n)$.

Now we give two examples as applications of the above upper bounds,
through which we obtain some idea about the order of teleportation
depth in comparison with the usual circuit depth.

\subsubsection{Teleportation depth of the $n$-qubit QFT}

\begin{figure}[htbp]
\includegraphics[width=3.40in]{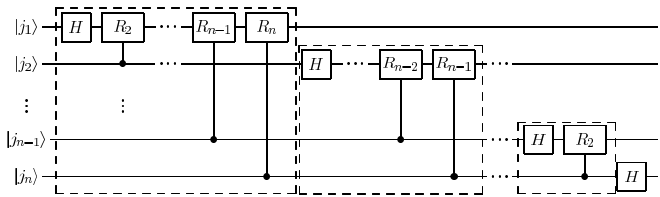}
\caption{Circuit for $n$-qubit Quantum Fourier Transform}\label{QFT}
\end{figure}

The first example is the $n$-qubit Quantum Fourier Transform (QFT)
circuit, as shown in FIG. \ref{QFT}. $R_k$ denotes the unitary
transformation $R_k=\text{diag}(1,e^{2\pi i/2^k})$. The circuit
depth of $n$-qubit QFT goes as $n^2$ and we will soon find that the
teleportation depth of this circuit is of the same order.

Each block of gates within a single dashed box (Hadamard plus
controlled $z$-rotations on the $k^{th}$ qubit)is a semi-Clifford
$(n-k+1)$-qubit $\mathcal{C}_{n-k+2}$ gate, $k = 1, \dots, n-1$ and
can be teleported directly using the one-bit scheme. Therefore the
whole circuit can be teleported piece by piece by one-bit
teleportation. Note that
\begin{eqnarray}
T(n,k=n+1) &\leq& T_1(n,k=n+1)\\&=&
2^n\left(1-(1-\frac{1}{2^n})^{n-1}\right)\\
&\sim& n-1
\end{eqnarray}
for large $n$. Actually,  numerical data shows that even when $n$ is
small, $T_1(n,k=n+1)\sim n-1$ is almost also true.

Therefore, the teleportation depth of the $n$-qubit QFT is
upper-bounded by
\begin{eqnarray}
\sum_{j=2}^{n} T(j,k=j+1)&\leq& \sum_{j=2}^{n} T_1(j,k=j+1)\\
& \leq & \sum_{j=1}^n (j-1)\nonumber\\
&=&\frac{1}{2}n(n-1)\sim\mathcal{O}(n^2).
\end{eqnarray}

Numerial calculation shows that $\sum_{j=2}^{n} T_1(j,k=j+1)$ is
almost $\frac{1}{2}n(n-1)-1$.

Note the probability for the teleportation process to terminate is
$1$ for teleporting an $n$-qubit $\mathcal{C}_{k = n+1}$ gate
$n-1=k-2$ times. This means that the upper bound we got for this
block teleportation scheme of QFT is just slightly lower than
naively assuming that we need $k-2$ teleportation steps to teleport
a $\mathcal{C}_k$ gate. The reason we do not benefit from the
avarage is that for QFT, $k$ is generally comparable with $n$.

\subsubsection{Uniformly Controlled rotation}

\begin{figure}[htbp]
\includegraphics[width=3.00in]{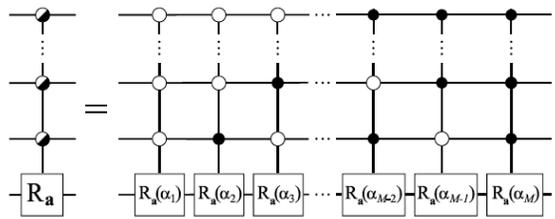}
\caption{Definition of the $n-1$-fold uniformly controlled rotation
of a qubit about the axis $\vec{a}$}\label{fig3}
\end{figure}

Now we consider another example, the uniformly controlled rotations,
which are widely used in analyzing the circuit complexity of an
arbitrary $n$-qubit quantum gate \cite{Mott} \cite{Vivek}. This
circuit in general needs $2^{n+2}-4n-4$ CNOT gates and $2^{n+2}-5$
one-qubit elementary rotations to implement. For complexity analysis
of this circuit see for example \cite{Mott}.

The teleportation depth of this rotation is in general upper bounded
by $2^n$. However, if each $(n-1)$-qubit-controlled gate is in
$\mathcal{C}_k$, we might expect to do better. For instance, when
$k=cn$, for any positive constant $n$, the teleportation depth
scales as $cn$, i.e. linear in $n$. Moreover, if $k\sim P(n)$, where
$P(n)$ is a polynomial in $n$, then the teleportation depth scales
as $P(n)$.

\subsection{Teleportation depth beyond semi-Clifford $\mathcal{C}_k$ gates}

Now recall our series of examples of non-semi-Clifford
$\mathcal{C}_k$ gates given in FIG. \ref{fig4}. We know that if
$V_k\in\mathcal{C}_k$, then $W_k\in\mathcal{C}_{k+1}$. And the group
$W_k\mathcal{P}_3W_k^{\dagger}$ does not contain a maximally abelian
subgroup of $\mathcal{P}_3$, i.e. $W_k\in\mathcal{C}_{k+1}$ is not
directly one-bit teleportable.

Therefore, we know that there are some $W_k$ gates in the
$\mathcal{C}_k$ hierarchy which can only be teleported directly by
the standard two-bit teleportation scheme. Using this scheme, we can
calculate another upper bound for teleportation depth, which we
denote as $T_2(n,k)$.

\begin{definition}
$T_2$ is the average total number of teleportation steps needed to
teleport separately each $\mathcal{C}_k$ component of a quantum
circuit using two-bit teleportation scheme, if such a decomposition
is possible.

More specifically, $T_2(n,k)$ is the average number of teleportation
steps needed to teleport an $n$-qubit $\mathcal{C}_k$ gate directly
(i.e. without decomposition) using the two-bit teleportation scheme.
\end{definition}

For a general $n$-qubit $\mathcal{C}_k$ gate, $T_2(n,k)$ can be
calculated by replacing $p$ with $\frac{1}{4^n}$ in Eq.
(\ref{depthck})
\begin{eqnarray}
T_2(n,k) &=& p\sum\limits_{s=1}^{k-3} s(1-p)^{s-1}+(k-2)(1-p)^{k-3}\nonumber\\
&=&4^n\left(1-(1-\frac{1}{4^n})^{k-2}\right)\label{depthck2}
\end{eqnarray}which then converges to $4^n$ when $k\rightarrow\infty$.

One may guess that in general to teleport $W_k$ directly using the
two-bit scheme will give a lower bound for teleportation depth than
to teleport the Toffoli gate and $V_k=diag(1,e^{i\pi/2^{k-1}})$
separately using the one-bit scheme. Surprisingly, this is not
generally true.

When $V_k\in\mathcal{C}_3$, this is indeed true. Teleporting $W_k$
directly gives a bound of $T_2(3,4)=1.875$, which is less than
$T_1(3,4)=2$, i.e. the bound given by teleporting the Toffoli gate
and $V_k$ separately with the one-bit scheme.

However, when $k\rightarrow\infty$, teleporting $W_k$ directly gives
a bound of $T_2(3,4)=5.25$, which is greater than $T_1(3,4)=3$, i.e.
the bound given by teleporting the Toffoli gate and $V_k$
separately.

This means that there exists a critical value $k$ that determines
which way is more efficient for teleporting $W_k$, directly or
separately.

Note if $V_k\in\mathcal{C}_k$, we also have
$W_k^{\dagger}\in\mathcal{C}_{k+1}$. Calculating the bounds of
teleportation depth for $W_k^{\dagger}$ shows a similar behavior as
that of $W_k$, however of a slightly different value. For instance,
when $V_k\in\mathcal{C}_3$, teleporting $W_k^{\dagger}$ directly
gives a bound of $1.5$, which is less than 2, the bound given by
teleporting separately. However when $k\rightarrow\infty$,
teleporting $W_k^{\dagger}$ directly gives a bound of $5.5$, but
teleporting separately gives only a bound of $3$.

Up to now, our discussion is entirely based on the $\mathcal{C}_k$
hierarchy. To summarize the capacity of $\mathcal{C}_k$ for
fault-tolerant quantum computation and provide basis for comparison
with non-$\mathcal{C}_k$ schemes discussed below, we introduce
another notion of $T_k$.

\begin{definition}
$T_k$ is the minimum number of total teleportation steps needed to
teleport separately each $\mathcal{C}_k$ component of a quantum
circuit using either one-bit or two-bit teleportation scheme.
\end{definition}

$T_k$ is defined in a way that represents the maximum capacity of
teleportation based on $\mathcal{C}_k$ hierarchy. In general $T_1
\geq T_k$, $T_2 \geq T_k$. To understand exactly how they compare
for a given circuit, a full characterization of the structure of
$\mathcal{C}_k$ is necessary. Here based on the structure theorems
given in Section III, we gave a simple example where $T_1$ or $T_2$
could be strictly larger than $T_k$. The next question to ask is
then whether we can go beyond $\mathcal{C}_k$ and this will be
discussed in the following section.

\subsection{Teleportation beyond $\mathcal{C}_k$}

In the definition of teleportation depth, we require that $A$ be
decomposed into a set of $\mathcal{C}_{\infty}$ gates. This is due
to the fact that $\mathcal{C}_{\infty}$ are the only gates that we
know so far how to perform fault-tolerantly by teleportation. In
general, if we do not require the decomposition to be in
$\mathcal{C}_{\infty}$, then we might get a better upper bound on
teleportation depth than the one defined previously, i.e. there
might exist upper bound $T^*$ of teleportation depth that is
strictly less than $T_k$. We give two such examples below. We leave
open the problem of how to implement teleportations fault-tolerantly
for a general $n$-qubit gate.
\\

\textbf{Example 1} For a general one-qubit gate $U$, we know that
$U$ can be decomposed into three $\mathcal{C}_{\infty}$ gates, each
of which has $T_1 < 2$. Hence through the decomposition we can bound
its total teleportation depth by $6$.

However, to teleport $U$ directly without decomposition via two-bit
teleportation gives a bound of $T_2 < 4^1 = 4$ less than $T_k$.
\\

\begin{figure}[htbp]
\includegraphics[width=1.00in]{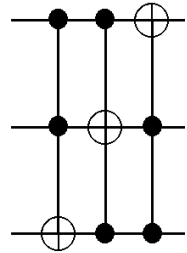}
\caption{The $R_{c3}$ gate--Three Toffoli gates in
series}\label{fig5}
\end{figure}

\textbf{Example 2} Consider a classical reversible circuit given in
FIG. \ref{fig5}. We denote this series of three Toffoli gates as
$R_{c3}$.

\begin{figure}[htbp]
\includegraphics[width=3.50in]{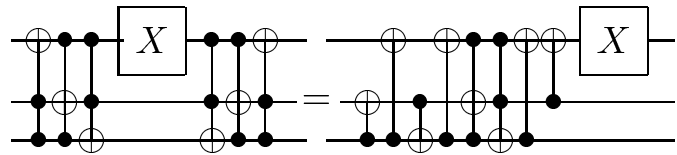}
\caption{Conjugating $X_1$ by $R_{c3}$}\label{fig6}
\end{figure}

This gate $R_{c3}$ is not in $\mathcal{C}_k$ hierarchy as can be
shown below:

Suppose that $R_{c3} \in \mathcal{C}_k$ is at certain level of the
hierarchy, $R_{c3} X_1 R_{c3}^{\dagger}$ must be a gate in
$\mathcal{C}_{k-1}$. Calculating explicitly as in FIG. \ref{fig6} we
have

The non-Clifford part of the right hand side of the equation is a
series of two Toffoli gates, and we denote it as $R_{c2}$. Due to
Proposition 3, $R_{c2}$ is also in $\mathcal{C}_{k-1}$.

\begin{figure}[htbp]
\includegraphics[width=3.00in]{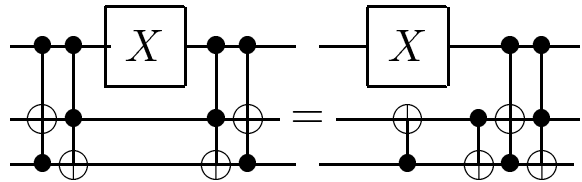}
\caption{Conjugating $X_1$ by $R_{c2}$}\label{fig7}
\end{figure}

As shown in FIG. \ref{fig7}, conjugating $X_1$ by $R_{c2}$ results in
$L R_{c2}^{\dagger}$, where $L$ is a Clifford operation. However, by
exchanging the second and third qubit in FIG. \ref{fig7}, we find
that $R_{c2}^{\dagger} X_1 R_{c2} = L' R_{c2}$, i.e. conjugating
$X_1$ by $R_{c2}^{\dagger}$ gives back $R_{c2}$. Therefore, $R_{c2}$
cannot be in the $\mathcal{C}_k$ hierarchy and we can conclude that
$R_{c3}$ is not a $\mathcal{C}_k$ gate either. $\square$

If we leave aside the problem of how to teleport gates beyond
$\mathcal{C}_k$ fault-tolerantly, we can teleport $R_{c3}$ directly
and obtain an upper bound of $2.75$, which is less than $T_k = 3$,
the bound given by teleporting the three Toffoli gates separately.

\section{Conclusion and Discussion}

In this paper we address the following questions: what is the
capacity of the teleportation scheme in practical implementation of
fault-tolerant quantum computation and what is the most efficient
way to make use of the teleportation protocol. To answer these
questions we first notice that  one-bit and two-bit teleportation
schemes require different resources to implement and are of
different capabilities. To understand what kind of gates can be
teleported fault-tolerantly with these two schemes respectively, we
study the structure of $\mathcal{C}_k$ hierarchy and its
relationship with semi-Clifford operations.  We show for $n=1,2$,
all the $\mathcal{C}_k$ gates are semi-Clifford operations, which is
also true for $\{n=3,k=3\}$. However, this is no longer true for
parameters $\{n>2,k>3\}$. Based on the counterexamples we
constructed for $\{n=3,k \geq 3\}$, we conjecture that all
$\mathcal{C}_3$ gates are semi-Clifford and all $\mathcal{C}_k$
gates are generalized semi-Clifford.

Such an understanding of the $\mathcal{C}_k$ structure has great
implications on the optimal design of fault-tolerant architectures.
While all $\mathcal{C}_k$ gates can be teleported fault-tolerantly,
the semi-Clifford subset of it requires less resources to implement
than others. To quantify this notion of gate complexity in
fault-tolerant quantum computation based on the $\mathcal{C}_k$
hierarchy, we introduce a measure called the teleportation depth
$T$, which characterizes how many teleportation steps are necessary,
on average, to implement a given gate.  Using different
teleportation schemes, we can give different upper bounds on $T$,
for example $T_1$, $T_2$ and $T_k$. General assumption was that $T_1
= T_2 = T_k = T$. However we showed in this work that, surprisingly
for certain series of gates $T_1$ could be strictly greater than
$T_k$ and $T_k$ could also be strictly greater than $T$.

The ultimate understanding of the structure of $\mathcal{C}_k$ will
provide a clearer clue on how to teleport circuits most efficiently.
To achieve this goal, some results from other branches of
mathematics might be helpful. It is noted that the Barnes-Wall
lattices, whose isometry group is a subgroup of index 2 in the real
Clifford group, have been extensively studied and recently their
involutions have been classified \cite{Griess}. It is our hope that
the $\mathcal{C}_3$ structure might be further understood once we
have a better understanding of the Clifford group.

For $n=1$, we fully characterize the structure of $\mathcal{C}_k$ by
further study on the diagonal gates in $\mathcal{C}_k$, which form a
group. It is interesting to note some evidence that $\mathcal{C}_k$
gates might be the only non-Clifford gates which could be
transversally implemented on a stabilizer code \cite{ZCC}. We also
fully characterize the structure of $\mathcal{C}_3$ for $n=3$, but
this seems not directly related to allowable transversal
non-Clifford gates on stabilizer codes. It is shown that those
transversal non-Clifford gates are allowed only if they are
generalized semi-Clifford \cite{Xie}, therefore we might expect some
generalized semi-Clifford $\mathcal{C}_k$ gates transversally
implementable on some stabilizer codes. We believe such kind of
exploration on the relationship between transversally implementable
gates and teleportable gates will shed some light on further
understanding of practical implementation of fault-tolerant
architectures.

\section*{Acknowledgments}

We thank Daniel Gottesman, Debbie Leung, and Carlos Mochon for comments.

\section*{Appendix A: Single qubit $\mathcal{C}_k$ gates}

\subsection*{1. Single qubit gates with eigenvalues $\pm 1$}

In this section we discuss what kind of single qubit unitary gates
could have eigenvalues $\pm 1$ apart from an overall phase factor,
i.e. if $\lambda_+,\lambda_-$ denote the two eigenvalues of a single
qubit unitary $U$, then what is the condition under which
$\lambda_++\lambda_-=0$. This information is useful since only the
unitary of this kind can be transformed into elements in Pauli group
under conjugation, i.e. there exits a unitary operator $R$, such
that $RAR^{\dag}=e^{i\theta}U$, where $A\in \mathcal{C}_{1}$. We'll
see that those kind of unitary has very restricted form which is
given by the following proposition.

\begin{proposition}:
The single qubit unitary gates which have eigenvalues $\pm 1$ apart
from an overall phase factor could only be of the following two
forms:
\begin{center}
$\Gamma_1(\varphi)=\left[
\begin{array}{cc}
0 & 1 \\
e^{i\varphi} & 0
\end{array}
\right]$
\end{center}
or
\begin{center}
$\Gamma_2(\phi,\xi)=\left[
\begin{array}{cc}
\cos{\phi}& \sin{\phi}e^{i\xi} \\
\sin{\phi}e^{-i\xi} & -\cos{\phi}
\end{array}
\right]$
\end{center}
\end{proposition}
\textbf{Proof}: We begin to prove this proposition by writing down a
general form of single qubit unitary gate as the following:
\begin{equation}
\Gamma=\left[
\begin{array}{cc}
\cos{\phi}e^{i\theta} & \sin{\phi}e^{i\xi} \\
\sin{\phi}e^{-i\xi} & -\cos{\phi}e^{-i\theta}
\end{array}
\right]
\end{equation}
Direct calculation gives
\begin{align}
\lambda_{\pm}&=
\frac{1}{2}\cos{\phi}e^{i\theta}-\frac{1}{2}\cos{\phi}e^{-i\theta}\nonumber\\
&\pm\frac{1}{2} e^{-i\theta}(\cos{\phi}^2 e^{4i\theta}-2\cos{\phi}^2
e^{2i\theta} +\cos{\phi}^2+4 e^{2i\theta})^{1/2}
\end{align}
Therefore $\lambda_++\lambda_-=0$ gives
\begin{equation}
\cos{\phi}\sin{\theta}=0
\end{equation}
If $\cos{\phi}=0$, the unitary must adopt the form of
$\Gamma_1(\varphi)$; if $\sin{\theta}=0$, then apart from an overall
phase, we can simply choose $\theta=0$ which leads to the form of
$\Gamma_2(\phi,\xi)$.$\square$

Note $\Gamma_1$ could be viewed as a special situation of $\Gamma_2$
for the case $\cos{\phi}=0$. However, we list $\Gamma_1$ separately
for future convenience.

\subsection*{2. Gate series associated with $\Gamma_1(\varphi)$ and
$\Gamma_2(\phi,\xi)$}

In this section we investigate the gate seises associated with
$\Gamma_1(\varphi)$ and $\Gamma_2(\phi,\xi)$. It is obvious that if
$\Gamma_1(\varphi), \Gamma_2(\phi,\xi) \in \mathcal{C}_{k}$, then
the unitary $U(\varphi)$ whose columns are the eigenvectors of
$\Gamma_1(\varphi)$ or $\Gamma_2(\phi,\xi)$ might be in
$\mathcal{C}_{k+1}$, given that
$U(\varphi)ZU(\varphi)^{\dag}=\Gamma_1(\varphi)$.

For $\Gamma_1(\varphi)$, the two normalized eigenvectors can be
chosen as
\begin{align}
|\Gamma_1(\varphi)\rangle_+&=\frac{1}{\sqrt{2}}(|0\rangle+e^{i\varphi/2}|1\rangle)\nonumber\\
|\Gamma_1(\varphi)\rangle_-&=\frac{1}{\sqrt{2}}(|0\rangle-e^{i\varphi/2}|1\rangle)
\end{align}
we now want a unitary whose columns is are eigenvectors of
$\Gamma_1(\varphi)$ apart from an overall factor of each
eigenvector, i.e.
\begin{align}
U(\varphi,\alpha)&=(e^{i\alpha}|\Gamma_1(\varphi)\rangle_+,
|\Gamma_1(\varphi)\rangle_-)\nonumber\\
&=\frac{1}{\sqrt{2}}\left[
\begin{array}{cc}
e^{i\alpha} & 1 \\
e^{i\alpha}e^{i\varphi/2} & -e^{i\varphi/2}
\end{array}
\right].
\end{align}
If $U(\varphi,\alpha)\in \mathcal{C}_{k+1}$, then
$U'=L_1U(\varphi,\alpha)L_2$ is also in $\mathcal{C}_{k+1}$. What is
important for us is to find $U'$ which is either of the form
$\Gamma_1$ or $\Gamma_2$, then from its eigenvectors we can generate
gates in $\mathcal{C}_{k+1}$. It is noticed that if we choose
$\alpha=0$, then
\begin{align}
U(\varphi,0)&=(|\Gamma_1(\varphi)\rangle_+,
|\Gamma_1(\varphi)\rangle_-)\nonumber\\
&=\frac{1}{\sqrt{2}}\left[
\begin{array}{cc}
1 & 1 \\
e^{i\varphi/2} & -e^{i\varphi/2}
\end{array}
\right],
\end{align}
and
\begin{align}
U(\varphi,0)HX=\left[
\begin{array}{cc}
0 & 1 \\
e^{i\varphi/2} & 0
\end{array}
\right]=\Gamma_1(\varphi/2).
\end{align}
Later we will show that for all the allowed value of $\alpha$, there
exist $L_1,L_2 \in \mathcal{C}_2$, such that
$L_1U(\varphi,0)L_2=U(\varphi,\alpha)$, so it is sufficient to
consider the case of $\alpha=0$.

Therefore we get a set of unitary given by
\begin{equation}
V_k(\varphi)=\Gamma_1(\varphi/2^{k}),
\end{equation}
if $\Gamma_1(\varphi) \in \mathcal{C}_2$ then
$\Gamma_1(\varphi/2^{k})$ could be in $\mathcal{C}_k$. We already
know that $\Gamma_(\pi/2)$ is in $\mathcal{C}_2$, then we have
\begin{equation}
V_k=\Gamma_1(2\pi/2^{k})
\end{equation}
is in $\mathcal{C}_k$.

Note
\begin{equation}
S_kX=V_k,
\end{equation}
and we already know that $S_k \in \mathcal{C}_k$. Therefore by
deriving $V_k$ we get nothing new due to proposition 1.

Now we come to the $\Gamma_2(\phi,\xi)$ case. Similarly, we begin
from the two normalized eigenvectors of $\Gamma_2(\phi,\xi)$, which
can be chosen as
\begin{eqnarray}
|\Gamma_2(\phi,\xi)\rangle_+&=&\frac{1}{\sqrt{2}}(\cos{\frac{\phi}{2}}|0\rangle+\sin{\frac{\phi}{2}}e^{-i\xi}|1\rangle)\nonumber\\
|\Gamma_2(\phi,\xi)\rangle_-&=&\frac{1}{\sqrt{2}}(\sin{\frac{\phi}{2}}e^{i\xi}|0\rangle-\cos{\frac{\phi}{2}}|1\rangle)
\end{eqnarray}
we now construct a unitary whose columns are eigenvectors of
$\Gamma_2(\varphi)$ apart from an overall factor of each
eigenvector, i.e.
\begin{align}
U(\phi,\xi,\beta)&=(e^{i\beta}|\Gamma_2(\phi,\xi)\rangle_+,
|\Gamma_2(\phi,\xi)\rangle_-)\nonumber\\
&=\frac{1}{\sqrt{2}}\left[
\begin{array}{cc}
e^{i\beta}\cos{\frac{\phi}{2}} & \sin{\frac{\phi}{2}}e^{i\xi} \\
e^{i\beta}\sin{\frac{\phi}{2}}e^{-i\xi} & -\cos{\frac{\phi}{2}}
\end{array}
\right].
\end{align}
If $U(\phi,\xi,\beta)\in \mathcal{C}_{k+1}$, then
$U'=L_1U(\phi,\xi,\beta)L_2$ is also in $\mathcal{C}_{k+1}$. It is
noticed that if we choose $\beta=0$, then
\begin{align}
U(\phi,\xi,0)&=(|\Gamma_2(\phi,\xi)\rangle_+,
|\Gamma_2(\phi,\xi)\rangle_-)\nonumber\\
&=\frac{1}{\sqrt{2}}\left[
\begin{array}{cc}
\cos{\frac{\phi}{2}} & \sin{\frac{\phi}{2}}e^{i\xi} \\
\sin{\frac{\phi}{2}}e^{-i\xi} & -\cos{\frac{\phi}{2}}
\end{array}
\right].
\end{align}
Also later we will show that for all the allowed value of $\alpha$,
there exist $L_1,L_2 \in \mathcal{C}_2$, such that
$L_1U(\phi,\xi,0)L_2=U(\phi,\xi,\beta)$, so it is sufficient to
consider the case of $\beta=0$.

Therefore we get a set of unitary given by
\begin{equation}
W_k(\phi,\xi)=\Gamma_2(\phi/2^{k-1},\xi),
\end{equation}
if $\Gamma_2(\phi,\xi) \in \mathcal{C}_2$ then
$\Gamma_1(\phi/2^{k-1},\xi)$ could be in $\mathcal{C}_k$. We already
know that only for $\Gamma_2(\pi/4,0)$ is in $\mathcal{C}_2$, then
we have
\begin{equation}
W_k=\Gamma_2(\pi/2^{k},0)
\end{equation}
is in $\mathcal{C}_k$.

Note for other possible values of $\phi$ and $\xi$, it is
straightforward to show that there exist $L_1,L_2 \in
\mathcal{C}_2$, such that
$L_1\Gamma_2(\pi/4,0)L_2=\Gamma_2(\phi,\xi)$, so it is sufficient to
consider the case of $\phi=\pi/4$ and $\xi=0$.

Note
\begin{equation}
HPW_kPX \sim S_k,
\end{equation}
where $\sim$ means up to an overall phase, and we already know that
$S_k \in \mathcal{C}_k$. Therefore again by deriving $W_k$ we get
nothing new due to proposition 1.

\subsection*{3. Gates in $\mathcal{C}_k \setminus \mathcal{C}_{k-1}$ for single qubit}

We conclude this section by presenting the following proposition,
which gives the structure of Gates in $\mathcal{C}_k \setminus
\mathcal{C}_{k-1}$ for single qubit.

\begin{proposition}
The set $\mathcal{C}_k \setminus \mathcal{C}_{k-1}$ for single qubit
is given by
\begin{equation}
L_1S_kL_2\in \mathcal{C}_k
\end{equation}
where $L_1,L_2\in \mathcal{C}_2$, $k\geq 2$.
\end{proposition}
\textbf{Proof}: We almost reached the proof of this proposition by
considering the results in subsections A and B. The only left we
need to clarify is

1. What happens when $\mathcal{C}_k$ is diagonal, which can not be
directly obtained by considering the eigenvectors of $V_{k-1}$ and
$W_{k-1}$. The answer is already known, since $S_k$ is the only
diagonal gate in $\mathcal{C}_k \setminus \mathcal{C}_{k-1}$.

2. The values of $\alpha$ and $\beta$. This can be answered by
noting the fact the equations
\begin{eqnarray}
UZU^{\dag}&=&G_1\nonumber\\
UXU^{\dag}&=&G_2
\end{eqnarray}
with $G_1,G_2$ known totally determines $U$ up to an overall phase.
Let's start from
\begin{equation}
U(\varphi,\alpha)=\frac{1}{\sqrt{2}}\left[
\begin{array}{cc}
e^{i\alpha} & 1 \\
e^{i\alpha}e^{i\varphi/2} & -e^{i\varphi/2}
\end{array}
\right].
\end{equation}
Note $U(\varphi,\alpha)ZU(\varphi,\alpha)^{\dag}\sim
\Gamma_1(2\phi)$, and
\begin{align}
&U(\varphi,\alpha)XU(\varphi,\alpha)^{\dag}\nonumber\\
=&\frac{1}{\sqrt{2}}\left[
\begin{array}{cc}
\cos{\alpha} & \sin{\alpha}e^{-i2\varphi}\\
\sin{\alpha}e^{i2\varphi} & -\cos{\alpha}
\end{array}
\right].\square
\end{align}

\section*{Appendix B: Detailed analysis about $\mathcal{C}_3$}

\subsection*{1. Notations}

Let's first define some notations.

Recall $\mathcal{P}_n$ is the Pauli group for $n$ qubit with order
$4^{n+1}$. Now let $\widetilde{\mathcal{P}}_n$ be the quotient group
$\mathcal{P}_n/Z(\mathcal{P}_n)$ with order $4^n$.

Let $\mathcal{C}_2(n)$ denote the Clifford group for $n$ qubit.
Define the quotient group
$\widetilde{\mathcal{C}}_2(n)=\mathcal{C}_2(n)/Z(\mathcal{C}_2(n))$.
Since $\widetilde{\mathcal{P}}_n$ is a normal subgroup of
$\widetilde{\mathcal{C}}_2(n)$, we could further define a quotient
group
$\widehat{\mathcal{C}}_2(n)=\widetilde{\mathcal{C}}_2(n)/\widetilde{\mathcal{P}}_n\cong
Sp(2n,2)$. Note $Sp(2,2)\cong S_3$ and $Sp(4,2)\cong S_6$. Denote
the set $\mathcal{K}(n)=\{A|A\in Sp(2n,2), A^2=1\}$, i.e.
$\mathcal{K}(n)$ are the set of all involutions of the symplectic
group $Sp(2n,2)$.

Denote the order of maximal Abelian subgroup of $\mathcal{K}(n)$ by
$a(n)$. Hence $a(1)=2,a(2)=8,a(n)\leq 2^{\frac{n(n+1)}{2}}$
\cite{Barry}.

Define the set $\mathcal{M}(n)=\{U|U\in
\widetilde{\mathcal{C}}_2(n)\setminus\widetilde{\mathcal{P}}_n\cup\{I\}\}$.

Now recall the definition for $\mathcal{C}_k(n)$:
\begin{equation}
\mathcal{C}_k(n)=\{U|UP_nU^{\dag}\in \mathcal{C}_{k-1}(n)\}
\end{equation}

For any $n$-qubit $U \in \mathcal{C}_k(n)$, the group $G_U(n)$ is
defined by $G_U(n)=U\widetilde{\mathcal{P}}_nU^{\dag}$.

Define the set $\mathcal{R}_k(n)=\{U|U\in
\widetilde{\mathcal{C}}_k(n),W^{\dag}=W,Tr(W)=0\}$.

And the set $\mathcal{F}_k(n)=\{U|U\in
\widetilde{\mathcal{C}}_k(n)$,$U$ is diagonal$\}$.

Denote the group generated by $\{A_i\}_{i=1}^{n}$ by $\langle
\{A_i\}_{i=1}^{n}\rangle$ for any set of operators $A_i$.

\subsection*{2. Some facts for calculating $\mathcal{C}_3$ structure}

We state some simple facts about $\mathcal{C}_3$ structure which we
use to verify Theorem 3 numerically.

\begin{fact}
We could always choose $G_U(n)\subset\mathcal{R}_{k-1}(n)$ for any
$U$ in $\mathcal{C}_k(n)$.
\end{fact}

Because we can always choose Hermitian and trace zero elements in
${\mathcal{P}}_n$ as the representative element for each element in
$\widetilde{\mathcal{P}}_n$.

\begin{fact}
If all $n-1$-qubit $\mathcal{C}_k$ gates are semi-Clifford, and if
$G_U(n)\supset \langle \{B_i\}_{i=1}^{n}\rangle$, where $B_i\in
\widetilde{\mathcal{P}}_n$ and $B_i\neq B_j, B_iB_j\neq B_k$ for
$i\neq j\neq k$, then $G_U(n)\cap \widetilde{\mathcal{P}}_n \subset
K_Z(n)$.
\end{fact}

Because if $\langle \{B_i\}_{i=1}^{n}\rangle\neq K_Z(n)$, then
$U(n)$ could be reduced to $U(1)\otimes U(n-1)$ via Clifford
operation.

\begin{fact}
If $A,B\in \mathcal{M}(n)\cap \mathcal{R}_2(n)$, and $A,B$
correspond to the same element in $\widehat{\mathcal{C}}_2(n)$, then
$AB\in \mathcal{P}_n$.
\end{fact}

Because if $A,B$ correspond to the same element in
$\widehat{\mathcal{C}}_2(n)$, then there exists $\alpha\in
\widetilde{\mathcal{P}}_n$ such that $A=\alpha B$.

\begin{fact}
For any $n$-qubit $\mathcal{C}_3$ gate $U$, if
$G_U(n)\supseteq\langle \{Z_i\}_{i=1}^{m}\rangle$, where $m\leq n$,
then the quotient group $G_U(n)/\langle \{Z_i\}_{i=1}^{m}\rangle\in
\mathcal{K}(n)$ is Abelian.
\end{fact}

For any $n$-qubit $\mathcal{C}_3$ gate $U$, if
$G_U(n)\supseteq\langle \{Z_i\}_{i=1}^{m}\rangle$, where $m\leq n$,
then the quotient group $G_U(n)/\langle \{Z_i\}_{i=1}^{m}\rangle\in
\mathcal{K}(n)$ is Abelian. Because elements of $G_U(n)\in
\widetilde{\mathcal{C}}_2(n)$ are either commute or anticommute, the
corresponding elements in $\widehat{\mathcal{C}}_2(n)$ should
commute.

\subsection*{3. $n=1$ case}

Since $Sp(2,2)\cong S_3$, $a(1)=2<4$. Hence $G_U(2)\cap
\widetilde{\mathcal{P}}_2$ contains at least one element in
$\widetilde{\mathcal{P}}_1$, i.e. $G_U(1)\cap
\widetilde{\mathcal{P}}_1 \supseteq K_Z(1)$ holds for any single
qubit $\mathcal{C}_3$ gate.

Furthermore, it is noted that any $U\in\mathcal{R}_k(1)$ can be
parameterized by
\begin{center}
$U(\theta,\varphi)=\left[
\begin{array}{cc}
\cos{\theta} & \sin{\theta}e^{i\varphi} \\
\sin{\theta}e^{-i\varphi} & -\cos{\theta}
\end{array}
\right]$,
\end{center}
and starting from elements in $\mathcal{R}_2(1)$ and calculate their
eigenvectors, we understand that $\varphi$ can only be of the values
$0,\frac{\pi}{2},\pi,\frac{3\pi}{2}$ for $\cos{\theta}\neq 0$. This
directly leads to the fact that the conjecture is true for any $k$
when $n=1$. See ck.pdf for the details of this.

\subsection*{4. $n=2$ case}

Since $Sp(4,2)\cong S_6$, $a(2)=8<16$. Hence $G_U(2)\cap
\widetilde{\mathcal{P}}_2$ contains at least one element in
$\widetilde{\mathcal{P}}_2$. However, this is not enough to claim
$G_U(2)\cap \widetilde{\mathcal{P}}_2$ holds for any two-qubit $U$.
We need to examine the structure of $G_U(2)\cap
\widetilde{\mathcal{P}}_2$ in more detail.

Consider the maximal Abelian subgroup in $\mathcal{K}(2)$ of order
$8$, and its corresponding elements in
$\widetilde{\mathcal{C}}_2(n)$, direct calculation shows it does not
contain a subgroup of structure $\widetilde{\mathcal{P}}_1\times
Z_2$. Hence we need to further consider Abelian subgroup in
$\mathcal{K}(2)$ of order $4$. Due to lemma 3, we result in
$G_U(2)\cap \widetilde{\mathcal{P}}_2 \supseteq K_Z(2)$ holds for
any two-qubit $\mathcal{C}_3$ gate.

Then using Lemma 1 and 2, we could calculate $\mathcal{C}_4(2)$
numerically. The result then shows that all the $\mathcal{C}_3(2)$
gates are semi-Clifford.

\subsection*{5. $n=3$ case}

Since a(3)=64, and direct calculation of this group shows that not
all the elements could be in $\mathcal{R}_2(3)$, hence $G_U(3)\cap
\widetilde{\mathcal{P}}_3$ contains at least one element in
$\widetilde{\mathcal{P}}_3$. Again, this is not enough to claim
$G_U(3)\cap \widetilde{\mathcal{P}}_3 \supseteq K_Z(3)$ holds for
any three-qubit $U$. We need to examine the structure of $G_U(3)\cap
\widetilde{\mathcal{P}}_3$ in more detail to dig out 2 more elements
in $\widetilde{\mathcal{P}}_3$.

Using Facts 1, 2 and 3, we could calculate $\mathcal{C}_3(3)$
numerically. The result shows that the conjecture is also true in
this case. See next subsection for more about $\mathcal{C}_3(3)$.

\subsection*{6. Diagonal gates in $\mathcal{C}_3$}

Define a diagonal Matrix  $A$ by $A_{jk}=\delta_{jk}e^{i\theta_j}$,
where $j=1,...,N,\ N=2^n$ , for $n$-qubit case.

We now prove the following

\begin{lemma}
If  $A\in\mathcal{C}_3$, if we choose $A_{11}=1$, then
$A_{jj}=e^{im_j\pi/4}$  for any $j\neq 1$ , where $m_j$  are some
integers.
\end{lemma}
\textbf{Proof}: We first prove for $j=N$ . Note we choose $A_{11}=1$
to get rid of the overall phase of  $A$. Denote  $A'=X^{\otimes
n}AX^{\otimes n}A^{\dagger}$, and  $A''=X^{\otimes n}A'X^{\otimes
n}A'^{\dagger}$. Note  $A'$, $A''$ are also diagonal. Since
$A\in\mathcal{C}_3$,  $A''$ must be in Pauli apart from an overall
phase. And we also have  $A''_{11}=e^{2i\theta_N}$,
$A''_{NN}=e^{-2i\theta_N}$. Hence we must have
$\frac{A''_{11}}{A''_{NN}}=e^{4i\theta_N}=\pm 1$ , i.e.
$\theta=\frac{m_N\pi}{4}$  for some integer  $m_N$.

For $j\neq N$ , there always exists a Clifford group operation which
keeps $\ket{j}$ invariant but maps
$\ket{1}\leftrightarrow\ket{N+1-j}$. Hence the above procedure
applies to any  $j\neq N$. $\square$

Note the similar idea applies to the diagonal $\mathcal{C}_k$ gates,
i.e. if $A\in\mathcal{C}_k$ , if we choose $A_{11}=1$ ,
$A_{jj}=e^{im_j\pi/2^{k-1}}$ for any $j\neq 1$, where $m_j$  are
some integers.

Now we consider some concrete gates:

\begin{proposition}
For  $n=3$,  the three qubit diagonal $\mathcal{C}_3$  gates are
given by a group generated by  $\pi/8$ gate, control-phase gate and
control-control-Z gate.
\end{proposition}

\textbf{Proof}: The proof is directly given by numerical
calculation, based on Lemma 1.$\square$

\end{document}